\documentclass[12pt]{iopart}
\usepackage{graphicx}
\usepackage{amsbsy}
\usepackage{color}

\begin{document}
\title{Probing Lorentz Invariance at EeV Energy } 
\author{Reetanjali Moharana and Nayantara Gupta}
\address{ Department of Physics, 
Indian Institute of Technology Bombay, Powai, Mumbai 400076, INDIA}
\eads{\mailto{reetanjali@iitb.ac.in}}
\begin{abstract}
Pierre Auger experiment has detected at least a couple of cosmic ray events above energy 60 EeV from the direction of the radio-galaxy Centaurus A. Assuming those events are from Centaurus A, we have calculated the numbers of neutral cosmic ray events from this source for small values of the degree of violation in Lorentz invariance. Our results show that a comparison of our calculated numbers of events with the observed number of events at EeV energy from the direction of the source can probe extremely low value of the degree of this violation.  
\end{abstract}

\maketitle
\section{Introduction}

The cosmic ray data from Akeno Giant Air Shower Array (AGASA) \cite{takeda}, Fly's Eye \cite{abu}, HiRes \cite{hires} and Pierre Auger \cite{Roulet} are helping us to understand the physics of the energetic cosmic ray particles coming from  the outer space. The important questions to be addressed are, what are their origin, mechanisms of production, composition and how they propagate through the interstellar medium. Due to the low flux of cosmic rays at ultrahigh energies $(>10^{17})$eV large scale detectors have to operate for many years to collect significant amount of signals from the ultrahigh energy universe.  
The composition of the ultrahigh energy cosmic rays is quite unknown. So far it has been determined using the hadronic interaction models and extrapolating the values of the particle interaction cross sections measured at low energies to ultrahigh energies. Novel methods of determining the composition of ultra high energy cosmic rays is a topic active research \cite{lem} . The highest energy cosmic ray event of energy $3.2\times10^{20}$eV was detected by Fly's Eye experiment \cite{Bird}. HiRes experiment, the upgraded version of 
Fly's Eye experiment, studied the highest energy cosmic rays in the energy range of $2\times10^{17}$eV to over $10^{20}$eV using the atmospheric fluorescence
 technique.
They reported the presence of GZK (Greisen-Zatsepin-Kuzmin) \cite{gre} suppression in the spectrum which is due to the interaction of very high energy cosmic ray protons and nucleons with cosmic microwave background (CMB) photons.

The Pierre Auger Observatory (PAO) is designed to study cosmic rays with energies more than $1$EeV. With its two sites in the two hemispheres this observatory is going to cover the entire sky. The northern site is going to be built in south east Colorado, USA. The southern site located in Mendoza, Argentina  has 1600 water Cherenkov surface detector stations covering 3000 $km^2$ and 24  fluorescence telescopes to record air shower cascades produced by cosmic rays.
Recently they have published results based on data taken between 1 January 2004 to 31 August 2007 \cite{auger1}. 
Cosmic ray protons of energy more than 57EeV are expected to be deflected by only a few degrees due to the Galactic and intergalactic magnetic fields during their propagation. 27 events have been detected with energies more than $57$EeV \cite{auger2}. Out of these events 20 correlate with at least one of the 442 AGN (Active Galactic Nuclei) at distances less than $75$Mpc. Their results are consistent with a GZK suppression in the cosmic ray spectrum.
Their correlation studies with extragalactic sources imply that the cosmic ray flux is not isotropic. Either the AGN or other astrophysical objects with similar spatial distribution are generating the extreme energy cosmic rays. Two events correlate with radio galaxy Cen A  within less than $3^{\circ}$ and several lie in the vicinity of its radio lobe near the super galactic plane. 15 events can be correlated with Seyfert galaxies among the closest AGN. More observational and theoretical investigations are necessary to confirm the sites of extreme energy cosmic ray production.
 At lower energy the deflection of cosmic ray protons are expected to be much more. As a result it is difficult to trace their origin.                                                         

Inside the cosmic accelerators both protons and neutrons are expected to be present, as the shock accelerated relativistic protons will produce neutrons in various interactions like $pp$ and $p\gamma$. Neutrons have a short lifetime and are not expected to reach the earth from distances of the order of 1 Mpc.
However, a small violation in the Lorentz invariance (LI) can lead to the stability of neutrons above a certain energy \cite{glash,dub}.
Also, protons become unstable and may decay to neutrons above a certain energy 
in this case. The prospect of exploring the possible minute violation in LI with energetic gamma ray and cosmic ray data has been studied in great detail by many physicists earlier \cite{am1,am2,steck1,jac1,jac2,jac3,jac4,gag,kol,steck2,uri,gal,steck3,xio}.       
Even extremely tiny violations in LI can change the physics of 
cosmic rays drastically \cite{wolf}.
In the present work we have calculated the expected number of neutron events 
from the closest radio-galaxy Centaurus A (Cen A) in the EeV energy range for different degrees of violation in the LI. It is shown that by comparing the observed event rate with our results it is possible to probe an extremely small value of this violation.   

\section {Violation in Lorentz Invariance (VLI)}
We briefly mention about the formalism of VLI used in the present work. The details are discussed in the original paper \cite{glash}. The authors of this paper assumed different maximum attainable velocities (MAVs) $c_a$ for different particles. The dispersion relation of the form $E^2=c_a^2p^2+m_a^2c_a^4$ describes a particle of type a moving freely in the preferred frame. For the neutron decay ($n \rightarrow p+e^-+\bar{\nu_e}$) 
 scenario the different MAVs for neutron, proton, electron and neutrino are $c_n$, $c_p$, $c_e$, $c_\nu$ respectively. Considering $c_p=c_e=c_\nu=c<c_n$ ($c$ is the speed of light) and using relativistic kinematics it is shown in \cite{glash} that neutron decay is forbidden above energy $E_1$, whose expression is given by
\begin{equation}
E_{1}= \sqrt{\frac{{m_n}^2-\left(m_p+m_e\right)^2}{c_p^2-c_n^2}}\simeq2.7\times 10^{19}\left[\frac{10^{-24}}{\delta}\right]^\frac{1}{2}ev
\end{equation}
Similarly protons become unstable ($p \rightarrow n+e^++\nu_e$) above energy $E_2$ 
\begin{equation}
E_2\simeq \sqrt{\frac{{m_n}^2-\left(m_p-m_e\right)^2}{c_p^2-c_n^2}}\simeq4.1\times 10^{19}\left[\frac{10^{-24}}{\delta}\right]^\frac{1}{2}ev
\end{equation}
In the above expressions $\delta=c_p-c_n$ denotes the extent of VLI. A small 
VLI can give rise to stable neutron events as neutrons do not decay and protons may decay to stable neutrons above energies $E_1$ and $E_2$ respectively. 
\section{Cosmic Ray Source and Spectrum}
 Cen A (NGC 5128) is the most promising source for UHECRs detected by PAO
. High flux of $\gamma $-rays with energy $>100$ GeV have been detected from Cen A \cite{gri,ste,all,cla}. Previously H.E.S.S \cite{aha} experiment detected gamma rays above $190$ GeV from this source. The flux 
 had an upper limit of $5.5\times 10^{-12}cm^{-2}s^{-1}$. This source was also observed earlier by EGRET \cite{sre} above 100MeV. 
Extrapolating the spectrum measured by EGRET with spectral index 2.4 above 190GeV one obtains photon flux comparable to that measured by H.E.S.S. experiment.
More observations have been carried out on Cen A in the recent past.
  H.E.S.S experiment has collected $\gamma$-ray data from Cen A above $\sim250$ GeV \cite{dis}. The data can be fitted with photon index $2.7 \pm 0.5_{stat}\pm 0.2_{sys}$. Fermi Gamma Ray Space Telescope has measured the flux $F\simeq 2.3\times 10^{-7}$ $ph$ $cm^{-2}$ $s^{-1}$ from this source above 100MeV. It is twice the flux measured by EGRET \cite{sre}.
In the past few years the high energy particle emission from Cen A has been investigated in various papers \cite{anc,cuo,gupta,kac}. Here we have explored the possibility of probing VLI of the order of $10^{-24}$ at EeV energy with Cen A as a source of cosmic rays. Protons are expected to be shcok accelerated to very high energy inside this radio galaxy. Neutrons can be produced inside the source in proton proton or proton photon interactions. These particles decay in a short time. However, a small VLI can stop the neutrons from decaying. The stable neutrons would be detectable by PAO or other large scale cosmic ray detectors.   
 The neutron flux (number of neutrons $per \, m^2 \, sec\, eV $) can be expressed as a power law in energy
\begin{equation}
\frac {dN_n(E_n)} {dE_n}=A_n \, \epsilon_n {E_n}^{-\alpha_n} 
\end{equation}
We have assumed a fraction $\epsilon_n$ of the total proton flux is converted to neutron flux inside the source.
These neutrons have energy more than $E_1$ which is determined by the degree of violation $\delta$ in LI.
The shock accelerated protons may become unstable above a energy $E_2$ due to VLI and give rise to neutrons.
The neutron flux produced in this way is
\begin{equation}
\frac {dN_n(E_n)} {dE_n}=A_n \, (1-\epsilon_n) {E_n}^{-\alpha_n} 
\end{equation}

We have used the total integrated exposure \cite{cuo} to calculate the number of neutron events expected in PA detector. The integrated exposure of PAO is $\Xi= 9000$ $km^2$ yr sr during $1^{st}$ Jan 2004 to $31^{st}$ Aug 2007. For a point source exposure per steradian ($\Xi/\Omega_{60}$) has been used where $\Omega_{60}=\pi$ sr. The relative exposure due to the angle of declination $-47^{\circ}$ is $\omega_s\simeq0.64$ for Cen A. PAO has detected $2$ events above $60$ EeV from the direction of Cen A. We have normalised the cosmic ray flux from the source using these two events. With different values of cosmic ray spectral index $\alpha_n$ and neutron production efficiency $\epsilon_n$ the expected numbers of neutron events above the threshold energy $E_1$ have been calculated.
The cosmic ray flux is not significantly attenuated during propagation due to interactions of neutrons with the photon background as Cen A is only 3.8 Mpc away.    
Charged cosmic ray particles will be deflected in all directions by the Galactic and extragalactic magnetic fields. These magnetic fields are not known to us very well. One may try to estimate the deflections of cosmic rays assuming the field remains unchanged over a certain length.  
The angle of deflection in the Galactic magnetic field of strength of the order of a few  $\mu$G  with a coherence length of order $\sim$ 1 kpc may be expressed
 as
\begin{equation}
\theta_1 \simeq 2.7^{\circ}\frac{60\textrm{ EeV}}{E}\left|\int\limits_{0}^{D}
\left(\frac{\textrm{d}\mathbf{x}}{\textrm{ kpc}} \times 
\frac{\mathbf{B}}{3\;\mu \textrm{G}}\right) 
\right|
\end{equation}
where the total distance traversed $D$ is about 20kpc \cite{stanev}. We have assumed $B=0.36 \mu$G in drawing Fig.1.
The root-mean-square deviation $\theta_2$ due to intergalactic magnetic field ($B_{rms}\sim 1nG$), coherence length $L_c\sim 1Mpc$ travelling a distance $D=3.8Mpc$ is shown in Fig.1. using the expression given below. 
\begin{equation}
\theta_{2} \approx 4^{\circ}\frac{60\textrm{ EeV}}{E}
\frac{B_{rms}}{10^{-9}\textrm{G}}\sqrt{\frac{D}{100\textrm{ Mpc}}}
\sqrt{\frac{L_{c}}{1\textrm{ Mpc}}}
\end{equation}
At EeV energy the protons will be deflected by large angles as one can see from
 Fig.1. As a result the proton background (from all sources including Cen A) to the neutron events from Cen A at
this energy is expected to be isotropic.  
\begin{figure}
  \begin{center}
    \begin{tabular}{cc}
      \includegraphics[width=8cm,angle=-90]{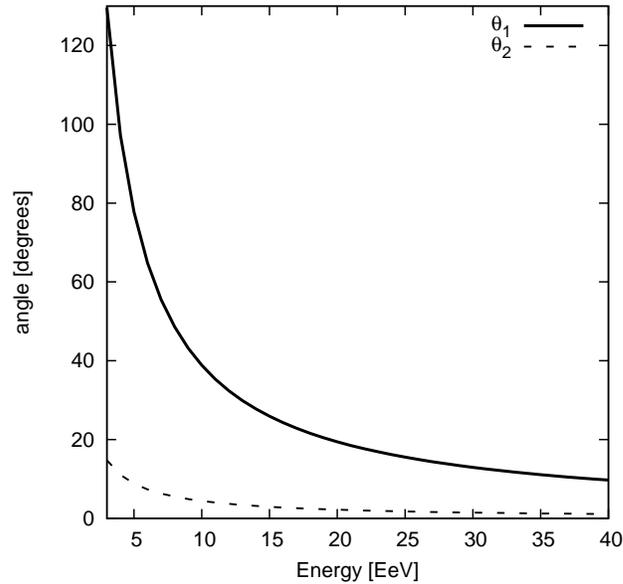}&
            \end{tabular}
\caption[...]{Deflection angles of protons in Galactic and extragalactic magnetic fields denoted by $\theta_1$ and $\theta_2$}
    \label{Figure:1}
  \end{center}
\end{figure}
The number of neutrons expected in PAO from the direction of Cen A above energy $E_1$ is 
\begin{equation} 
N_{1}=A_{n}\epsilon_n\frac{\Xi\,\omega_s}{\Omega_{60}}\displaystyle\int^\infty_{E_1}E_n^{-\alpha_n}dE_n
\end{equation}
Above energy $E_2$ the number of neutrons from decaying protons is expected to be 
\begin{equation} 
N_{2}=A_{n}(1-\epsilon_n)\frac{\Xi\,\omega_s}{\Omega_{60}}\displaystyle\int^\infty_{E_2}E_n^{-\alpha_n}dE_n
\end{equation}
The total numbers of neutron events $(N_1+N_2)$ from Cen A expected in PAO during $1$ January $2004$ to $31$ August 2007 have been plotted in Fig.2 . A variation in $\delta$ leads to variations in $E_1$ and $E_2$. Fig.3. shows the variation in $\delta$ with $E_1$.

\begin{figure}
  \begin{center}
    \begin{tabular}{cc}
      \includegraphics{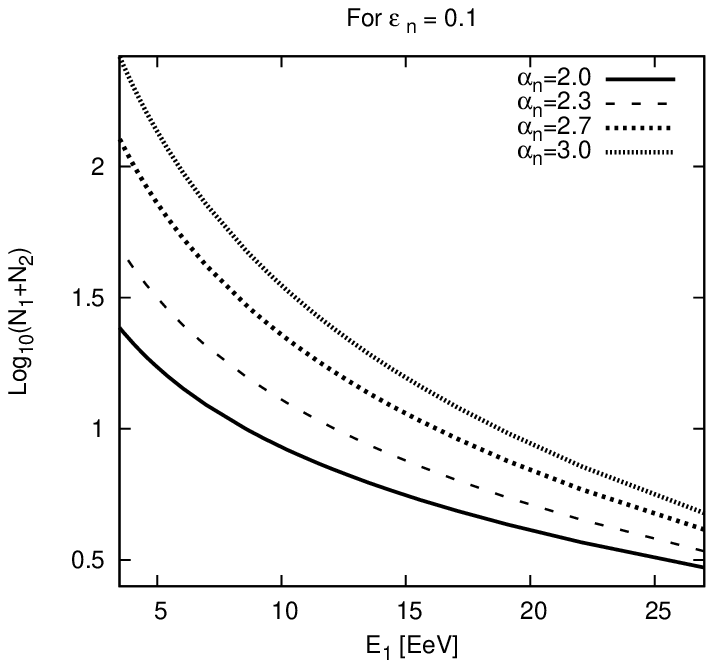}&
      \includegraphics{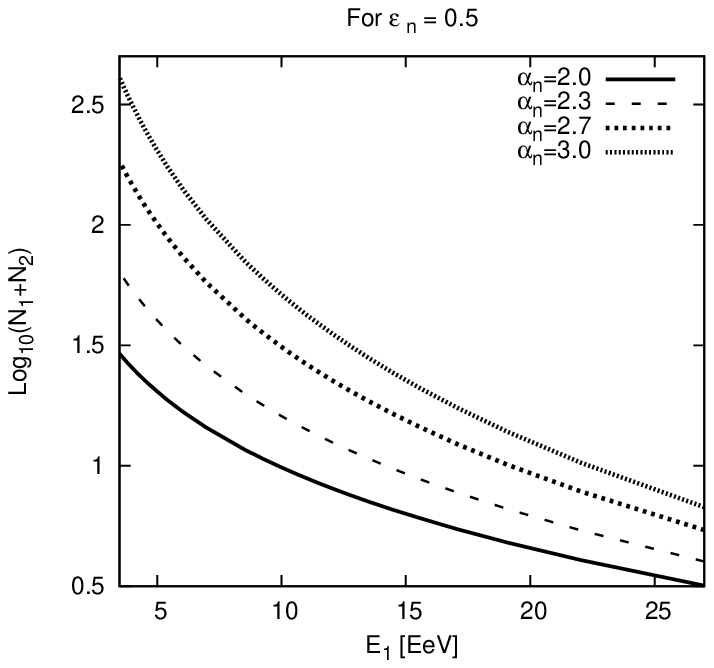} \\
      \end{tabular}
\caption[...]{Figure shows number of neutrons predicted from Cen A for different values of $\epsilon_n$ }
    \label{Figure:2}
  \end{center}
\end{figure}

\begin{figure}
  \begin{center}
    \begin{tabular}{cc}
      \includegraphics{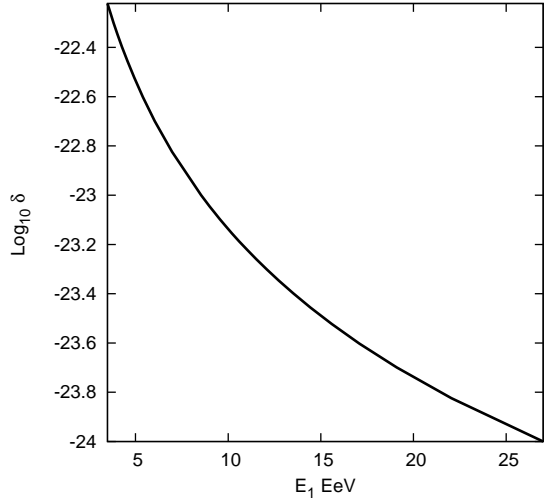}&
      \end{tabular}
\caption[...]{Figure shows variation in $\delta$ with $E_1$ }
    \label{Figure:3}
  \end{center}
\end{figure}
\section {Detectability of Neutron Events}
Fig.2. shows the number of neutron events expected above energy $E_1$ in PAO from Cen A for the integrated exposure $(9000/\pi) km^2$yr. Above 25EeV energy a few  neutron events are expected from this source. $E_1=25$EeV corresponds to 
$\delta=10^{-24}$. More neutron events can be collected for longer duration of observation. Hence, it is possible to probe very small values of $\delta$. In Fig.4 the background cosmic ray flux from the direction of the source within an area of $3^{\circ}\times3^{\circ}$ has been compared with the neutron flux from the source. The neutron flux is higher than the background above 5EeV, hence it is easy to detect whether there is any excess of events from the direction of Cen A. Here it is important to discuss about the various possibilities. The two events correlated within $3^{\circ}$ of Cen A may have a different origin. 
Our analysis does not depend on the type of the source.
 If the Galactic magnetic field in the vicinity of the sun is less than 0.24$\mu$gauss \cite{hug} such correlation of events is possible even with protons. The positional accuracy of PAO for UHECR events is less than $0.6^{\circ}$. The observed neutron events should be coming from within this angle from their origin.  
The anisotropy observed in the direction of Cen A extends to 
$20^ \circ$. Detection of a large number of neutron events within $1^{\circ}$ of Cen A can confirm whether there is a small VLI at EeV energy.
\begin{figure}
  \begin{center}
    \begin{tabular}{cc}
      \includegraphics{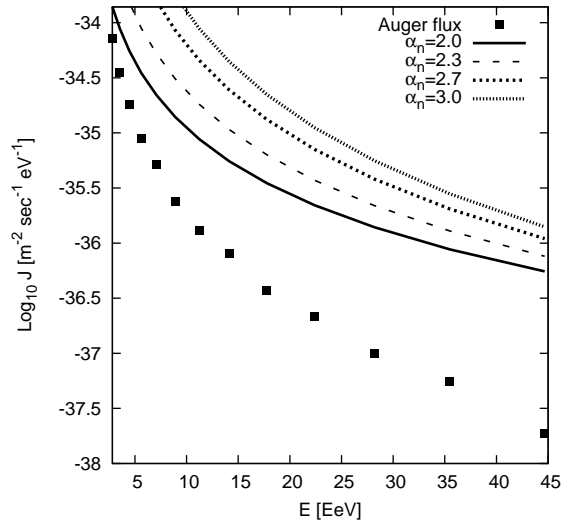}
      \end{tabular}
\caption[...]{The cosmic ray flux from PAO \cite{abr} within $3^{\circ}\times3^{\circ}$ of Cen A (denoted by square points) has been compared with neutron flux (different line styles used for different spectral indices) from this source}
    \label{Figure:4}
  \end{center}
\end{figure}

\section{Conclusion}
We have discussed about a possibility of probing small values of VLI at EeV energy using large scale cosmic ray detectors. The two events detected within $3^{\circ}$ of Cen A have been used to normalise the cosmic ray flux from the direction of Cen A. In future if more cosmic ray sources are identified close to us then they can be used to probe VLI in the same way. A large number of neutron events within $1^{\circ}$ of the source would be a signature of VLI at EeV energy.
\section{Acknowledgement}
We thank the anonymous referee for constructive comments.


\newpage
\begin{thebibliography}{99}
\bibitem{takeda} Takeda M. et al., 1998 Phys. Rev. Lett. {\bf 81} 1163; 
1999 ApJ {\bf 522} 225; see also http://www-akeno.icrr.u-tokyo.ac.jp/AGASA/
\bibitem{abu} Abu-Zayyad T. et al., 2004 Phys. Rev. Lett. {\bf 92} 151101 
\bibitem{hires}http://www.cosmic-ray.org/ 
\bibitem{Roulet} Roulet E. for the Pierre Auger Collaboration, 2008 J. Phys. Conf. Ser {\bf 136} 022051; see also http://www.auger.org/
\bibitem{lem} Lemoine M. and Waxman E.,arxiv:0907.1354
\bibitem{Bird} Bird D. J. et al., 1993 Phys. Rev. Lett. {\bf 71} 3401
\bibitem{gre} Greisen K., 1966 Phys. Rev. Lett. {\bf 16} 748; Zatsepin G. T. and Kuzmin V. A., 1966 Sov. Phys. JETP Lett. {\bf 4} 78
\bibitem{auger1} Pierre Auger Collaboration, 2007 Science {\bf 318} 938 
\bibitem{auger2} Pierre Auger Collaboration, 2008 Astropart. Phys. {\bf 29} 188 ; 2008 Erratum-ibid 30:45
\bibitem{glash} Coleman S. and Glashow S. L., 1999 Phys. Rev. D {\bf 59} 116008
\bibitem{dub} Dubovsky S. L. and Tinyakov P. G., 2002 Astropart. Phys. {\bf 18} 89
\bibitem{am1} Amelino-Camelia G. et al., 1998 Nature {\bf 393} 763 
\bibitem{am2} Amelino-Camelia G. and Piran T., 2001 Phys. Rev. D {\bf 64} 036005 
\bibitem{steck1} Stecker F. W. and Glashow S. L., 2001 Astropart. Phys. {\bf 16} 7; F. W. Stecker, 2003 Astropart. Phys. {\bf 20} 85 
\bibitem{jac1} Jacobson T. et al., 2002 Phys. Rev. D {\bf R 66} 081302
\bibitem{jac2} Jacobson T. et al., 2003 Nature {\bf 424} 1019 
\bibitem{jac3} Jacobson T. et al., 2004 Phys. Rev. Lett. {\bf 93} 021101
\bibitem{jac4} Jacobson T. et al., 2005 Lec.Notes in Phys. {\bf 669} 101 
Editors:Jurek Kowalski-Glikman, Giovanni Amelino-Camelia. 
\bibitem{gag} Gagnon O. and Moore G. D., 2004 Phys. Rev. D {\bf 70} 065002 
\bibitem{kol} Chisholm J. R. and Kolb E. W., 2004 Phys. Rev. D {\bf 69} 085001 
\bibitem{steck2} Stecker F. W. and Scully S. T., 2005 Astropart. Phys. {\bf 23} 203 
\bibitem{uri} Jacob U. and Piran T., 2007 Nature Phys. {\bf 3} 87 
\bibitem{gal} Galaverni M. and Sigl G., 2008 Phys. Rev. Lett. {\bf 100} 021102 
\bibitem{steck3} Scully S. T. and Stecker F. W., 2009 Astropart. Phys. {\bf 31} 220; Stecker F. W. and Scully S. T., arxiv:0906.1735
\bibitem{xio} Xiao-Jun Bi et al., 2009 Phys. Rev. D {\bf 79} 083015
\bibitem{wolf} Bietenholz W., arxiv:0806.3713
%\bibitem{Gur} Sergey Gureev et al.; http://arxiv.org/abs/0808.0481 
\bibitem{gri} Grindlay J. E. et al., 1975 Ap. J. {\bf 197} L9
\bibitem{ste} Steinle H., 1998 A \& A {\bf 330} 97
\bibitem{all} Allen W. H. et al., 1993a Astropart. Phys. {\bf 1} 269
\bibitem{cla} Clay R. W., Dawson B. R. and Meyhandan R., 1994 Astropart. Phys. {\bf 2} 347
\bibitem{aha} Aharonian F. et al., 2005 A \& A {\bf 441} 465
%\bibitem{kab} Kabuki S. et al., 2007 Ap. J. {\bf 668} 968
\bibitem{sre} Sreekumar P. et al., 1999 Astropart. Phys. {\bf 11} 221
\bibitem{dis} Aharonian F. et al, arxiv:0903.1582, accepted for publication in ApJ Letters
\bibitem{fermi} Abdo A. A. et al.,arxiv:0902.1559
\bibitem{anc} Anchordoqui L. A., Goldberg H. and Weiler T. J., 2001 Phys. Rev. Lett. {\bf 87} 081101
\bibitem{cuo} Cuoco A. and Hannestad S., 2008 Phys.Rev.D {\bf 78} 023007 
\bibitem{gupta} Gupta N., 2008 JCAP {\bf 06} 022
\bibitem{kac} Kachelrie{\ss} M. et al, arxiv:0805.2608 
\bibitem{stanev} Stanev T., 1997 Astrophys. J. {\bf 479} 290
\bibitem{hug} Hague, J. D. et al., Pierre Auger Collaboration, arxiv:0906.2347
\bibitem{abr} Pierre Auger Collaboration (J. Abraham et al.), 2008 Phys. Rev. Lett. {\bf 101} 061101 
\end{thebibliography}
\end{document}